# Chaotic advection in three-dimensional unsteady incompressible laminar flow

By JULYAN H. E. CARTWRIGHT[1], MARIO FEINGOLD[2], AND ORESTE PIRO[1]

[1]Departament de Física, Universitat de les Illes Balears, 07071 Palma de Mallorca, Spain.
julyan@hp1.uib.es & piro@hp1.uib.es

[2]Department of Physics, Ben-Gurion University, Beer-Sheva 84105, Israel.
mario@bgumail.bgu.ac.il

We discuss chaotic advection in three-dimensional unsteady incompressible laminar flow, and analyse in detail the most important novel advection phenomenon in these flows; the global dispersion of passive scalars in flows with two slow and one fast velocity components. We make a comprehensive study of the first model of an experimentally realizable flow to exhibit this *resonance-induced* dispersion: biaxial unsteady spherical Couette flow is a three-dimensional incompressible laminar flow with periodic time dependence derived analytically from the Navier–Stokes equations in the low-Reynolds-number limit.

chao-dyn/9504012  16 Aug 95

## 1. Introduction

There is a hierarchy in the dynamics of incompressible laminar† fluid flows. A flow can be one, two, or three-dimensional, and it may be steady (time independent) or unsteady (time dependent). The character of a laminar flow varies greatly depending on to which category in this hierarchy it belongs. Of course, a fluid possesses three spatial dimensions, and in this trivial sense a fluid flow is always a three-dimensional entity. However, in many instances the fluid velocity is zero in one or more coordinate directions, and here one can speak of a one or two-dimensional flow. Another possible way for a one or two-dimensional flow to arise — a generalization of the above — is that one or more of the three velocity components may be independent of the others. This category, which might be termed that of $(2+1)$-dimensional flows, comes between truly two-dimensional flows where the third velocity component is zero, and truly three-dimensional flows where all three velocity components are coupled. An example of $(2+1)$-dimensional flow‡ is spherical Couette flow; the steady flow at low Reynolds numbers between coaxially-rotating concentric spheres, where there is both a primary one-dimensional flow and a secondary two-dimensional flow mathematically completely independent of the primary flow, although physically induced by it. In this paper we take advantage of the spherical symmetry of the geometry of this system to couple the primary and secondary flows. The coupling produces a truly three-dimensional flow — biaxial unsteady spherical Couette flow — which we can model analytically as a solution of the Navier–Stokes equations. The purpose of this paper is to study in detail the dynamics of biaxial unsteady spherical Couette flow. We present results showing that a novel and important phenomenon of global resonance-induced particle dispersion takes place, and show that such resonance-induced dispersion is generic in three-dimensional unsteady incompressible laminar flows with two slow and one fast velocity components.

---

† We are using *laminar* to mean non-turbulent here — the (Eulerian) velocity at any given point in space is fixed, periodic, or quasiperiodic in time — following the generally established usage. It should be born in mind, however, that the laminae are not necessarily either topologically simple or fixed in time (Aref *et al.* 1989).

‡ Another flow that falls into this category is the eccentric helical annular mixer model of Kusch and Ottino (1989; 1992): regular duct flows in general are of this type.

　We are interested in particle paths (path lines) in laminar flow. Idealized particles that are small enough not to perturb the flow, but large enough not to engage in Brownian motion (i.e., not to diffuse), and that move with the flow itself are known as passive scalars. The same physics applies to scalar or vector quantities possessed by a fluid, such as temperature, concentration of a second fluid, or magnetic field. These quantities move with the flow — are passively advected — and so may be described by the Lagrangian representation of the fluid; their dynamics is that of a fluid element

$$\dot{x} = u_x(x,y,z,t), \qquad \dot{y} = u_y(x,y,z,t), \qquad \dot{z} = u_z(x,y,z,t), \qquad (1.1)$$

where $u_x$, $u_y$, and $u_z$ are the Cartesian components of the velocity field $\boldsymbol{u}(u_x, u_y, u_z)$. We shall be studying laminar flows where $\boldsymbol{u}$ is time periodic (unsteady) or time independent (steady), that are incompressible, that is to say $\nabla \cdot \boldsymbol{u} = 0$.

　Before embarking on our exploration of three-dimensional unsteady flows, we briefly recall the state of knowledge of advection in laminar flows at lower levels of the flow hierarchy, starting with the simplest case. One-dimensional incompressible laminar flows, such as Poiseuille flow or Couette flow, are encountered at the beginning of fluid-dynamics textbooks, precisely because their dynamics, whether steady or unsteady, is very simple: one dimensionality plus incompressibility means that the fluid velocity must be constant at any time in the direction of flow. In two dimensions, incompressibility of the flow implies that $\partial u_x/\partial x + \partial u_y/\partial y = 0$, which tells us that there exists an exact differential $\mathrm{d}\psi$ such that

$$\dot{x} = \frac{\partial \psi}{\partial y}, \qquad \dot{y} = -\frac{\partial \psi}{\partial x}. \qquad (1.2)$$

$\psi$ is known as the stream function, and we can immediately recognize these equations as forming a Hamiltonian dynamical system. The motion of a passive scalar in a two-dimensional incompressible flow can then be identified with the Hamiltonian dynamics of a point in phase space, with the stream function $\psi$ as the Hamiltonian. Accordingly, two-dimensional steady incompressible flows are, like one-degree-of-freedom time-independent Hamiltonians, integrable. Two-dimensional unsteady incompressible flows, on the other hand, have the dynamics of one-degree-of-freedom time-dependent Hamiltonians, which are generically nonintegrable. In the case where the time dependence is periodic, these have stroboscopic maps — mappings from one period to the next — which are area-preserving maps on the plane, and show the chaotic motion typical of these maps, with chaotic trajectories bounded by the invariant KAM tori predicted by the famous Kolmogorov–Arnold–Moser theorem. The apparition of chaotic stream lines in the Lagrangian picture of fluid dynamics has been termed chaotic advection, or Lagrangian turbulence† and is important not least because it greatly enhances transport in these flows compared with their steady counterparts. Both numerical (Aref 1984; Aref & Balachander 1986; Chien, Rising & Ottino 1986; Khakhar, Rising & Ottino 1986; Chaiken *et al.* 1987) and experimental (Chaiken *et al.* 1986; Solomon & Gollub 1988; Swanson & Ottino 1990) investigations of chaotic advection in two-dimensional unsteady incompressible laminar flows have been undertaken. Many of these investigations have used as a model the unsteady Couette flow between two eccentric cylinders, often termed unsteady journal-bearing flow, for which an analytical solution of the Navier–Stokes equations is available in the low-Reynolds-number limit. Unsteady eccentric Couette flow has proved to be an ideal experimental and numerical model for chaotic advection in two dimensions.

---

† We prefer the former terminology to the latter, since it is not clear what relationship there is between chaotic advection and turbulence.

The steady three-dimensional incompressible flow problem is similar to the unsteady case in two dimensions, since the return map to any section transverse to the flow is an area-preserving map on the plane. Moreover, a steady three-dimensional incompressible flow can be reduced to a one-degree-of-freedom time-dependent Hamiltonian system, except at points where the flow has a stagnation point (in the terminology of dynamical systems, a *fixed point*) (Arnold 1978; Bajer 1994; MacKay 1994). If there are no stagnation points and if the flow in one coordinate direction is never zero, this coordinate can be used as a fictitious time (Aref *et al.* 1989; MacKay 1994). Arnold (1965) first addressed chaos in three-dimensional steady incompressible inviscid fluid flow. He proved that if the vorticity $\boldsymbol{\omega} = \nabla \times \boldsymbol{u}$ is nowhere parallel to the velocity $\boldsymbol{u}$ then the flow is integrable. At the same time he conjectured that Beltrami flows where vorticity and velocity are everywhere parallel, $\boldsymbol{\omega} = \lambda \boldsymbol{u}$, would have stream lines of complex topology. Hénon (1966) took up this suggestion and examined the $\lambda = 1$ case, now known as ABC (Arnold–Beltrami–Childress) flow. The ABC flow, an inviscid flow — a solution of the three-dimensional Euler equation — was subsequently studied by Dombre *et al.* (1986), and has become the archetypal mathematical model for the study of chaos in three-dimensional steady incompressible laminar flow. Interesting chaotic Beltrami flows in a sphere — the spherical analogues of ABC flow — were analysed by Zheligovsky (1993). Other mathematical models of chaotic advection in three-dimensional steady Stokes flows within a spherical drop have been investigated by Bajer & Moffatt (1990), and Stone, Nadim & Strogatz (1991). Three-dimensional steady flows that are experimentally realizable have also been shown to exhibit chaotic advection. The partitioned-pipe mixer model of Khakhar, Frajione & Ottino (1987) has been experimentally studied by Kusch & Ottino (1992). Similarly, the twisted pipe model of Jones, Thomas & Aref (1989) has led to experimental investigations by Le Guer, Castelain & Peerhossaini (1995).

Three-dimensional unsteady incompressible laminar flow has an entirely different dynamics. We survey in §2 the types of chaotic advection to expect in the three-dimensional unsteady case using a classification of these unsteady volume-conserving flows by the number of fast and slow components of the velocity field. Concentrating in §3 on flows with two slow and one fast velocity components, in §3.1 we review spherical Couette flow, in order that in §3.2 we may go on to fulfil our aim in this paper, to explore in detail a model we (Cartwright, Feingold & Piro 1994a, 1995) recently introduced of an experimentally realizable flow that exhibits the most important novel advection phenomenon in three-dimensional flows; the global *resonance-induced* dispersion of passive scalars. Finally, in §4, we draw these different threads together and discuss the possibilities for experiments based on this and other unsteady three-dimensional flows.

## 2. Classifying three-dimensional unsteady flows with Liouvillian maps

A classical technique, due to Poincaré, for the study of a system of ordinary differential equations describing a dynamical system such as a fluid flow, is the reduction of the dimension of the system achieved by sectioning the flow in time or space to produce a map. The dynamical behaviour of this map can then be investigated much more easily than that of the original system, in the knowledge that its limit sets are simply related to the limit sets of the flow (Arrowsmith & Place 1990). The technical details of this procedure differ depending on whether the flow is autonomous (steady) or nonautonomous (unsteady) (Parker & Chua 1989), and the terminology is not completely fixed: here we take a Poincaré section to mean any such section, and a Poincaré map to be the dynamics on the section. We talk of a stroboscopic map when we refer to the Poincaré map of a nonautonomous system obtained by sampling the flow stroboscopically once per period. The Poincaré map of an autonomous system obtained by taking the intersections of the trajectory (path line) with a fixed plane we refer to as a return map.

Three-dimensional unsteady incompressible flows with periodic time dependence produce stroboscopic maps in three dimensions that preserve volume. These three-dimensional volume-preserving maps do not correspond to Hamiltonian systems, which have phase spaces of even dimensionality: Hamiltonians with two or three degrees of freedom produce volume-preserving stroboscopic maps of two or four dimensions respectively. Following Liouville's theorem, we have named volume-preserving maps of three dimensions, which dimensionally interpolate between the two cases above, *Liouvillian* maps (Feingold, Kadanoff & Piro 1987; Feingold, Kadanoff & Piro 1988a; Feingold, Kadanoff & Piro 1988b; Piro & Feingold 1988; Feingold, Kadanoff & Piro 1989; Cartwright, Feingold & Piro 1994b).

To study Liouvillian maps one can employ ideas from KAM theory and imagine perturbations of integrable action–angle models where the actions are fixed and the angles change at each iteration. A perturbation away from the integrable case then causes the actions to vary slowly. For two-dimensional area-preserving maps, or for any maps derived from Hamiltonian dynamical systems, the underlying symplectic structure provides the action–angle variables on which to build. Here with Liouvillian maps such symplectic structure is absent, so we must define our variables with respect to something else. Thus we use invariant tori to define the action–angle variables.

We can consider integrable, Liouvillian maps that possess some variables, actions $I$, that do not change from iteration to iteration, whereas others, angles $\theta$, vary linearly as the map is iterated†:

$$\boldsymbol{I}' = \boldsymbol{I}, \qquad \boldsymbol{\theta}' = \boldsymbol{\theta} + \boldsymbol{\omega}(\boldsymbol{I}). \tag{2.1}$$

We use the notation $\boldsymbol{L}_k^0$ to represent an integrable Liouvillian map which possesses a set of action–angle variables $\boldsymbol{I} \in \mathbb{R}^k$, $\boldsymbol{\theta} \in \mathbb{T}^{3-k}$. A $k$-action integrable Liouvillian map $\boldsymbol{L}_k^0$ has a family of $(3-k)$-dimensional invariant tori, each of which is transversed by the $3-k$ angle variables that rotate at each iteration by an amount given by the corresponding component of the $(3-k)$-dimensional vector $\boldsymbol{\omega}$. In other words, $\boldsymbol{L}_k^0$ maps consist of uniform translations on $(3-k)$-tori embedded in $\mathbb{T}^3$. We thus have four possibilities for these integrable maps:

$\boldsymbol{L}_0^0$. Zero actions; three angles. The frequency vector $\boldsymbol{\omega}$ is constant, and hence we have a uniform rotation on $\mathbb{T}^3$.

$\boldsymbol{L}_1^0$. One action; two angles. The motion takes place on two-tori defined by constant $I$, and a point on one of these tori is given by the two angles $\theta_1$ and $\theta_2$. Thus the motion on the tori is a uniform rotation with a frequency depending on the value of the action.

$\boldsymbol{L}_2^0$. Two actions; one angle. The two actions parametrize a family of invariant circles (i.e., one-tori), and the angle variable rotates with a frequency $\omega(I_1, I_2)$.

$\boldsymbol{L}_3^0$. Three actions; zero angles. Each point of space is parametrized by $(I_1, I_2, I_3)$ and is hence invariant; the map is the identity.

We consider now the effects of nonintegrable perturbations on $\boldsymbol{L}_k^0$ maps. The perturbations we wish to consider will act to couple the hitherto independent actions and angles whilst retaining the Liouvillian property of the map. Thyagaraja & Haas (1985) have worked out the most general form that such perturbations can take, which in general leads to an implicit map. Below we shall discuss only the subset of perturbations that leave the map in explicit form. We believe that these explicit Liouvillian maps capture the dynamics of Liouvillian maps in general, whilst being easier to work with than implicit maps.

We deal first with the two extreme cases $k = 0$ and $k = 3$. Results of Ruelle & Takens (1971) and Newhouse, Ruelle & Takens (1978) about perturbing quasiperiodic flows on a three-torus indicate that small perturbations of zero-action $\boldsymbol{L}_0^0$ maps will generically lead to completely chaotic behaviour‡. A small nonintegrable perturbation added to three-action $\boldsymbol{L}_3^0$ maps is a small perturbation

$$I_1' = I_1 + \varepsilon F_1(I_2, I_3), \qquad I_2' = I_2 + \varepsilon F_2(I_1', I_3), \qquad I_3' = I_3 + \varepsilon F_3(I_1', I_2'), \qquad (2.2)$$

of the identity. In the $\varepsilon \to 0$ limit we have a set of three autonomous ordinary differential equations, which takes us back to the case of three-dimensional steady flows. In other words, the perturbation represents the *Euler map* of the flow $\dot{\boldsymbol{I}} = \boldsymbol{F}(\boldsymbol{I})$; the map (2.2) is derived from this flow by applying the forwards and backwards Euler numerical integration methods to discretize it. The dynamics of Euler maps are closely related to the flows from which they are derived (Cartwright & Piro 1992).

---

† Notice that at this stage it is not important whether the frequency vector is made a function of the old actions $\omega(\boldsymbol{I})$ or of the iterated actions $\omega(\boldsymbol{I}')$. This will become important when we wish to perturb this map whilst retaining its Liouvillian nature: we construct the perturbation to make it automatically volume preserving.

For one-action $\boldsymbol{L}_1^0$ maps, it turns out that most of the invariant two-tori are preserved in a KAM-like manner under small nonintegrable volume-preserving perturbations. However, the resonant tori satisfying the equation

$$m\omega_1(I) + n\omega_2(I) = 2\pi k, \qquad (2.3)$$

where $k$, $m$, and $n$ are integers, are destroyed, and in their place appear a finite number of invariant circles, half of which are stable and the other half unstable, in the same way as invariant circles are destroyed and fixed points appear in planar area-preserving maps following the Poincaré–Birkhoff theorem. Further invariant tori surround the stable invariant circles, whilst chaos appears around the unstable ones. The preserved invariant tori act as boundaries for the chaotic layers, so a single trajectory cannot cover the whole phase space until the perturbation is large enough to destroy the final invariant torus. The whole situation is like that governed by the KAM theorem in planar area-preserving maps, except that the structures have one additional dimension; that is the invariant tori are two rather than one-dimensional, the island chains interspersed amongst them contain invariant circles, not fixed points, and the stable (elliptic) and unstable (hyperbolic) invariant circles are centres for further invariant tori and for chaos respectively (Feingold, Kadanoff & Piro 1988b; Cartwright, Feingold & Piro 1994b). The existence of a KAM-like theorem in this one-action case has been proven by Cheng & Sun (1990), although as Mezić & Wiggins (1994) point out, here in three dimensions we have no criterion similar to the irrationality of rotation numbers in two dimensions that enables one to predict in which order the remaining invariant tori are destroyed as the nonintegrable perturbation is increased.

With two-action $\boldsymbol{L}_2^0$ maps, on the other hand, small nonintegrable volume-preserving perturbations cause behaviour rather different to that detailed above. The integrable case turns out to be a singular limit; as soon as a nonintegrable perturbation is switched on, the two-parameter family of invariant circles coalesce into invariant two-tori. We can see why this should be so using an argument based on an adiabatic approximation. Consider the perturbed two-action map

$$I_1' = I_1 + \varepsilon F_1(I_2, \theta), \qquad I_2' = I_2 + \varepsilon F_2(I_1', \theta), \qquad \theta' = \theta + \omega(I_1', I_2'). \qquad (2.4)$$

Suppose that $\omega$ is irrational, then when $\varepsilon$ is small, the angle $\theta$ covers uniformly the entire interval $(0, 2\pi)$ before the actions $\boldsymbol{I}$ change significantly. The variation of $\boldsymbol{I}$ is thus sensitive only to the average of $\boldsymbol{F}$ over $\theta$. Therefore the actions now iterate as

$$I_1' = I_1 + \varepsilon \bar{F}_1(I_2), \qquad I_2' = I_2 + \varepsilon \bar{F}_2(I_1'). \qquad (2.5)$$

where the bar represents the $\theta$ average. Thus the dynamics of the actions $\boldsymbol{I}$ decouples from that of the angle $\theta$ for non-resonant $\omega$. We are left with an area-preserving map, which leads in the

‡ However, the effect of restricting the perturbations so that they retain volume conservation has not, as far as we know, been investigated.

limit $\varepsilon \to 0$ to the Hamiltonian system

$$\dot{I}_1 = \bar{F}_1(I_2) = \frac{\partial H}{\partial I_2}, \qquad \dot{I}_2 = \bar{F}_2(I_1) = -\frac{\partial H}{\partial I_1}. \tag{2.6}$$

We obtain the Hamiltonian

$$H(I_1, I_2) = \int_0^{I_2} \bar{F}_1(I) \mathrm{d}I - \int_0^{I_1} \bar{F}_2(I) \mathrm{d}I = \beta. \tag{2.7}$$

In other words, in the $\varepsilon \to 0$ limit, the action variables evolve slowly along the level curves of (2.7). Including the fast motion in the $\theta$ direction, we infer that the originally invariant circles parallel to the $\theta$ axis coalesce into invariant tori $\Sigma_\beta$ defined by the condition $H(I_1, I_2) = \beta$, the motion on which is fast in one direction and slow in the other. Resonances occur when the condition imposed above that $\omega(\boldsymbol{I})$ be irrational is not satisfied and we have

$$n\omega(I_1, I_2) = 2\pi k, \tag{2.8}$$

where $k$ and $n$ are integers. The resonant condition defines a set of sheets that typically intersect a continuous set of invariant tori, which break down locally at the intersections. The remainder of the invariant torus survives, to a given order in the perturbation expansion. Invariant tori are thus connected through resonances, meaning that, in contrast with the one-action case, a single trajectory can cover the whole of phase space (Piro & Feingold 1988; Feingold, Kadanoff & Piro 1989; Cartwright, Feingold & Piro 1994b). We term this characteristic behaviour of two-action Liouvillian maps *resonance-induced* dispersion.

## 3. Resonance-induced dispersion in two-action flows

As we have seen, two-action Liouvillian maps have particularly interesting behaviour, since there are no barriers to motion and a form of global dispersion is present. To investigate this resonance-induced dispersion further in a fluid dynamical context, in this section we construct a model of a realistic flow showing two-action properties.

### 3.1. *Spherical Couette flow*

Spherical Couette flow — steady flow at low Reynolds numbers between coaxially-rotating concentric spheres — is a natural generalization of the Couette flow between two rotating cylinders. In contrast with the latter, however, it possesses both a primary and a secondary flow, and it is dependent on the Reynolds number. References to experimental and numerical studies of spherical Couette flow are given by Koschmieder (1993).

Consider concentric spheres with fixed radii $R_1$ and $R_2$, and constant angular velocities about a common axis through their centres $\tilde{\Omega}_1$ and $\tilde{\Omega}_2$, as pictured in figure 1. In the spherical polar coordinate system of figure 1, with the radial coordinate rescaled as $r = \tilde{r}/R_2$, the steady flow of a viscous, incompressible fluid contained between the two spheres is assumed to be independent of the longitude $\phi$. We may then write the Navier–Stokes equations in terms of a stream function $\psi$ in the meridian plane, and a longitudinal velocity function $\Omega$

$$\frac{1}{Re} \mathrm{D}^2 \Omega = -\frac{\psi_r \Omega_\theta - \psi_\theta \Omega_r}{r^2 \sin^2 \theta}, \tag{3.1}$$

$$\frac{1}{Re} \mathrm{D}^4 \psi = \frac{2\Omega}{r^3 \sin^2 \theta} (\Omega_r r \cos\theta - \Omega_\theta \sin\theta) + \frac{2\mathrm{D}^2 \psi}{r^3 \sin^2 \theta} (\psi_r r \cos\theta - \psi_\theta \sin\theta)$$
$$- \frac{1}{r^2 \sin\theta} \left( \psi_r (\mathrm{D}^2 \psi)_\theta - \psi_\theta (\mathrm{D}^2 \psi)_r \right), \tag{3.2}$$

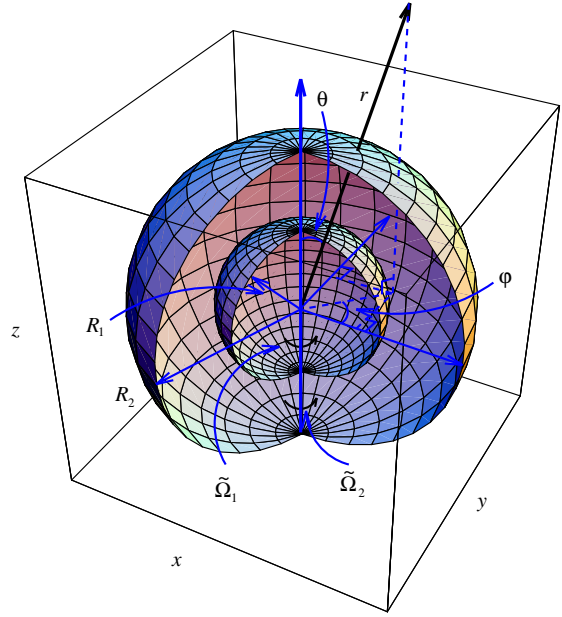

FIGURE 1. Two concentric spheres have radii $R_1$ and $R_2$, and constant angular velocities $\tilde{\Omega}_1$ and $\tilde{\Omega}_2$ about a common axis through their centres. The space between them is filled with an incompressible viscous fluid. We term the ratio $R_1$ to $R_2$, $\eta$, and the ratio $\tilde{\Omega}_1$ to $\tilde{\Omega}_2$, $\mu$. The radial coordinate is rescaled as $r = \tilde{r}/R_2$ in order that it be dimensionless.

where

$$\mathrm{D}^2 = \frac{\partial^2}{\partial r^2} + \frac{1}{r^2}\frac{\partial^2}{\partial \theta^2} - \frac{1}{r^2}\cot\theta\frac{\partial}{\partial \theta}, \tag{3.3}$$

and the subscripts associated with $\psi$ and $\Omega$ denote differentiation.

The Reynolds number is defined as $Re = \tilde{\Omega}_c R_2^2/\nu$, where $R_2$ is a characteristic length given by the radius of the outer sphere, $\tilde{\Omega}_c$ is a characteristic angular velocity, and $\nu$ is the kinematic viscosity. The characteristic angular velocity may be taken as that of the outer sphere $\tilde{\Omega}_2$, or that of the inner sphere $\tilde{\Omega}_1$, depending on which is the dominant influence on the flow in the case being considered. The boundary conditions are firstly that $\psi = \psi_r = 0$ on both spheres, when $r = \eta$ and $r = 1$. Secondly, with $\tilde{\Omega}$ as the physical angular velocity, and choosing $\tilde{\Omega}_2$ as the characteristic angular velocity, the dimensionless angular velocity $\omega = \tilde{\Omega}/\tilde{\Omega}_2 = \Omega/r^2 \sin^2\theta$ should be $\omega = \mu$ on the inner sphere and $\omega = 1$ on the outer. Hence the other two parameters, apart from the Reynolds number, are the radius ratio $\eta = R_1/R_2$, and the angular velocity ratio $\mu = \tilde{\Omega}_1/\tilde{\Omega}_2$.

The above problem can be solved analytically by a perturbation method in the Reynolds number. Pearson (1967) gave the solution to first order in the Reynolds number for the case $\mu = 0$, $\eta = 1/2$. Subsequently Munson & Joseph (1971) derived general results through to seventh order in the Reynolds number. Here we need only the first order approximation valid in the low-Reynolds-number limit $Re \lesssim 1$ (with the higher-order solutions one can obtain good approximations for Reynolds numbers in the hundreds). The solution to first order in the Reynolds

number is

$$\psi = Re\left(\frac{A_1}{r^2} + A_2 + A_3 r^3 + A_4 r^5 + \frac{a_2}{4}\left(\frac{a_2}{r} - a_1 r^2\right)\right) \sin^2\theta \cos\theta, \qquad (3.4)$$

$$\Omega = \left(a_1 r^2 + \frac{a_2}{r}\right) \sin^2\theta, \qquad (3.5)$$

where

$$a_1 = \frac{1 - \mu\eta^3}{1 - \eta^3}, \qquad a_2 = \frac{(\mu - 1)\eta^3}{1 - \eta^3}, \qquad (3.6)$$

and

$$A_1 = \Xi\eta^4\Big(\left(2\eta^5 - \eta^4 - 16\eta^3 - 20\eta^2 - 8\eta - 2\right)\mu \\ + 2\eta^5 + 8\eta^4 + 20\eta^3 + 16\eta^2 + \eta - 2\Big), \qquad (3.7)$$

$$A_2 = -\Xi\eta^2\Big(\left(10\eta^7 + 25\eta^6 + 40\eta^5 + 57\eta^4 + 48\eta^3 + 20\eta^2 + 8\eta + 2\right)\mu \\ - 2\eta^7 - 8\eta^6 - 20\eta^5 - 48\eta^4 - 57\eta^3 - 40\eta^2 - 25\eta - 10\Big), \qquad (3.8)$$

$$A_3 = -\Xi\Big(\left(6\eta^6 + 24\eta^5 + 35\eta^4 + 32\eta^3 + 15\eta^2 - 2\eta - 5\right)\mu\eta^2 \\ + 5\eta^6 + 2\eta^5 - 15\eta^4 - 32\eta^3 - 35\eta^2 - 24\eta - 6\Big), \qquad (3.9)$$

$$A_4 = \Xi\Big(\left(2\eta^4 + 8\eta^3 + 5\eta^2 - 2\eta - 3\right)\mu\eta^2 + 3\eta^4 + 2\eta^3 - 5\eta^2 - 8\eta - 2\Big), \qquad (3.10)$$

with

$$\Xi = \frac{(\mu - 1)\eta^3}{4(4\eta^6 + 16\eta^5 + 40\eta^4 + 55\eta^3 + 40\eta^2 + 16\eta + 4)(\eta - 1)^2(\eta^2 + \eta + 1)^2}. \qquad (3.11)$$

The dimensionless flow velocities are given by

$$u_r = \frac{1}{r^2 \sin\theta}\frac{\partial \psi}{\partial \theta} = \dot{r}, \qquad (3.12)$$

$$u_\theta = -\frac{1}{r \sin\theta}\frac{\partial \psi}{\partial r} = r\dot{\theta}, \qquad (3.13)$$

$$u_\phi = \frac{1}{r \sin\theta}\Omega = r \sin\theta\, \dot{\phi}, \qquad (3.14)$$

so that

$$\dot{r} = \frac{Re}{r^2}\left(\frac{A_1}{r^2} + A_2 + A_3 r^3 + A_4 r^5 + \frac{a_2}{4}\left(\frac{a_2}{r} - a_1 r^2\right)\right)(3\cos^2\theta - 1), \qquad (3.15)$$

$$\dot{\theta} = -\frac{Re}{r^2}\left(-\frac{2A_1}{r^3} + 3A_3 r^2 + 5A_4 r^4 + \frac{a_2}{4}\left(-\frac{a_2}{r^2} - 2a_1 r\right)\right) \sin\theta \cos\theta, \qquad (3.16)$$

$$\dot{\phi} = a_1 + \frac{a_2}{r^3}. \qquad (3.17)$$

The solution shows a primary flow $\Omega$ about the axis of rotation, together with a secondary flow given by $\psi$ in the meridian plane. In general, one of the two spheres dominates the flow pattern, generating a pair of vortices, one on either side of the equator, whose sense of rotation depends on the angular velocity ratio $\mu$. The spheres may corotate or counterrotate, and in the latter case ($\mu$ negative) there is a part of the $(\eta, \mu)$ parameter space where neither sphere is dominant and we observe two counterrotating vortices in each hemisphere (Yavorskaya *et al.* 1980), giving a total of four vortices. Figure 2 shows examples of both the two-vortex and the four-vortex patterns. Since all three velocity components depend only on $r$ and $\theta$, spherical Couette flow is in the category of flows we named $(2 + 1)$-dimensional in §1. The secondary flow stream function $\psi$ is a one-degree-of-freedom time-independent Hamiltonian of the type (1.2), so the steady flow is identifiable as an integrable Hamiltonian system. Now consider the flow in terms of the categorization we discussed in §2. Inserting typical values for the magnitudes of the variables and parameters (by which we mean typical of those we will use throughout the rest of this paper) into (3.15–3.17) shows that at the values of the Reynolds number $Re \leqslant 1$ for which (3.15–3.17) hold, the one-dimensional primary flow $\Omega$ governed by $\phi$ is hundreds to thousands of times faster than the two-dimensional secondary flow $\psi$ governed by $r$ and $\theta$. Thus we have two slow action variables $r$ and $\theta$, and one fast angle variable $\phi$; the flow falls into the two-action category.

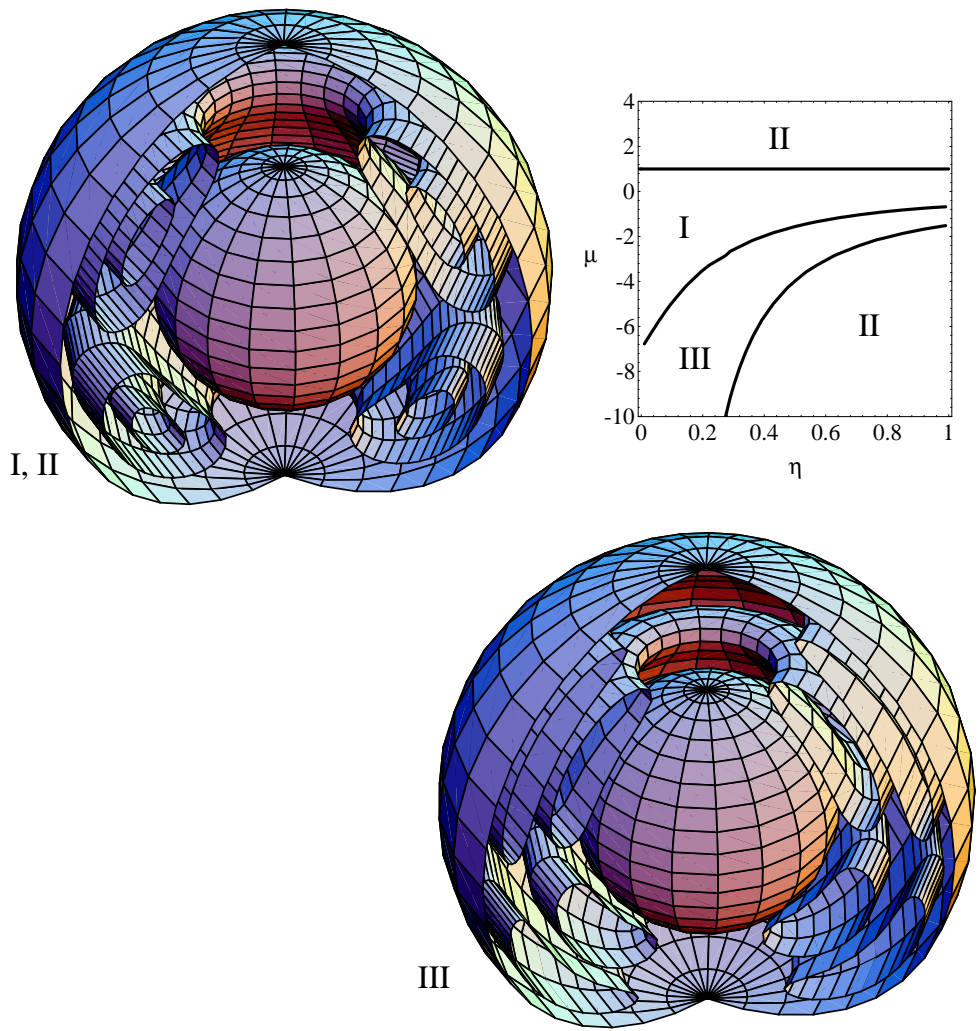

FIGURE 2. Shown here are the regions of parameter space for which the different secondary flow patterns between the spheres exist, and typical examples of the two-vortex and four-vortex flow are represented as cutaway pictures showing the spheres together with some level surfaces of the vortices. When $\mu = 1$ there is rigid body rotation and no secondary flow. Either side of this line two vortices are present. In region I the outer sphere is the dominant influence on the flow, and fluid is transported from the poles to the equator near the outer sphere, and back near the inner sphere. In the two disjoint parts of region II the inner sphere is the dominant influence, and fluid is transported in the opposite direction, from the equator to the poles near the outer sphere, and back near the inner sphere. In between, in the linguiform region III, neither sphere is the dominant influence and there are four vortices. These rotate such that fluid is transported to the equator near the inner and outer spheres, and to the poles at the separatrix between the two vortices of each hemisphere.

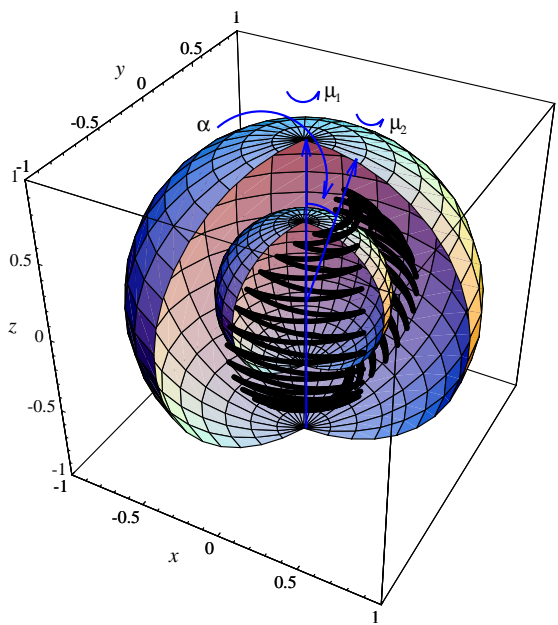

FIGURE 3. A trajectory in biaxial unsteady spherical Couette flow illustrates the piecewise (mathematically speaking, $C^0$) nature of the solution. The trajectory covers thirty periods. The radius ratio is $\eta = 0.5$, the angular velocity ratios are $\mu_1 = 0$ and $\mu_2 = -3$, and the semiperiods are $T_1 = T_2 = T/2 = 2$. The Reynolds number is $Re = 0.1$ and the axis separation is $\alpha = 10°$. The initial conditions are $r = 0.64$, $\theta = 60°$, and $\phi = 30°$.

### 3.2. *Biaxial unsteady spherical Couette flow*

The existence of a secondary flow in spherical Couette flow is crucial for our purposes, since it enables us to construct a fully three-dimensional flow simply by using two different axes of rotation for the spheres. Rotating periodically about first one axis, then the other acts to couple the longitude to the latitude and the radial coordinate, giving a three-dimensional flow that is also time dependent, figure 3. We call the time for which we rotate about the first axis, the first semiperiod, $T_1$, and that about the second axis, the second semiperiod, $T_2$, making a total forcing period of $T = T_1 + T_2$. For simplicity we shall set $T_1 = T_2 = T/2$, and we shall vary $T$ as necessary.

Biaxial unsteady spherical Couette flow is then a piecewise combination of the flow solutions about each axis. This piecewise combination of solutions is the same technique used to obtain the unsteady journal-bearing flow extensively investigated as a model for chaotic advection in two-dimensional unsteady flows. However, an important conceptual difference between spherical Couette flow and journal-bearing flow is that in the latter problem the solution depends linearly on the rotation rates of the inner and outer cylinders. Thus in journal-bearing flow, particle paths depend only on the boundary displacements. As long as the cylinders are set in motion alternately, the details of the modulation of the rotation rate are not important; provided the angular displacements of the boundaries are the same, so will be the path of a passive scalar. This is not true of spherical Couette flow, since the stream function is not a linear function of the rotation rate. Thus in the spheres case we have to decide the way in which to modulate the rotation. The two extremes are the discontinuous limit, and the quasistatic limit. In the discontinuous limit we change the rotation rate as discontinuously as possible, with instantaneous startup and shutoff of the rotation, as with a square wave. We then model the problem as a piecewise-linear combination of the two separate steady spherical Couette flow solutions. This is a reasonable approximation if the fluid inertia is negligible, so that the fluid stops moving as soon as the sphere stops rotating, which is why we require low-Reynolds-number flow. In the quasistatic limit, on the other hand, we change the rotation rate as slowly as possible, with smooth modulation of the rotation in time, as with a sine wave. The quasistatic approximation then consists of considering the resulting stream function to be given instantaneously by the steady stream function for the rotation rate at that instant. We have chosen to use the discontinuous approximation to model the unsteady flow here. We shall use radius ratio $\eta = 1/2$ throughout this paper, and we can elect to use different angular velocity ratios $\mu_1$ and $\mu_2$ in each semiperiod to alter the character of the flow solution.

The alternative to the above choices for modulating the rotation rates of the spheres would be to model the full developing flow; the so-called spin-up problem discussed by Pearson (1967). This makes the model far more complex, and we feel justified in ignoring the complication since the discrepancy is small at low Reynolds numbers. In the case of either unsteady journal-bearing flow or biaxial unsteady spherical Couette flow, one is neglecting transient effects such as the diffusion of momentum. This neglect requires that the acceleration terms in the Navier–Stokes equation

$$Re \left( Sr \frac{\partial \boldsymbol{u}}{\partial t} + \boldsymbol{u}.\nabla \boldsymbol{u} \right) = -\nabla p + \nabla^2 \boldsymbol{u} \tag{3.18}$$

be negligible, where $Re = VL/\nu$ is the Reynolds number as before, and $Sr = fL/V$ is the Strouhal number, with $L$ a characteristic length, $V$ a characteristic velocity, and $f$ a characteristic frequency; $\nu$ being the kinematic viscosity. To make the acceleration terms negligible, *Re* and *Sr* must be small. This last condition is violated by the discontinuous modulation of the rotation rate in time; at the discontinuities, $\partial \boldsymbol{u}/\partial t$ will be infinite. However, we argue that the effects of this violation will be slight at low Reynolds numbers (many of the studies of journal-bearing flow have been made with discontinuous changes in the cylinder rotation rates), and to place us squarely in the low Reynolds number regime we set $Re = 0.1$ throughout this paper. Because the dynamical phenomena we observe in this system are structurally stable, that is to say robust against small changes in the system, the small extra displacement suffered by a fluid particle not responding instantaneously when the flow stops and restarts will not lead to a qualitative change in its long term behaviour. This robustness is also a prerequisite for being able to observe these phenomena in a real fluid experiment.

To explore the dynamics of a three-dimensional unsteady flow, we may plot a stroboscopic map, which gives us a three-dimensional volume-preserving map like the Liouvillian maps of §2, but a three-dimensional view of a Liouvillian map often fails to give much information. In order to reveal the information contained within a three-dimensional map, we need to open it up to view. There are various methods to achieve this; we shall use two in this paper. The first is to slice the map open by taking a section through it. We select a thin slice and record the iterates that lie within it. There is a trade off between the thickness of the slice and the quality of the information that is obtained; a thicker slice has more points lying within it, but is fuzzier because of its three dimensionality, a thinner slice gives a sharper image but requires longer time integration of the system to have the same number of points in the slice. In the spheres case we use a slice between $-0.01 < y < 0.01$. The second technique we use is to project the map down onto a plane. Unlike the slice technique, which will always produce useful information, this technique is only useful if the Liouvillian map is near to being integrable. If we are in possession of the action and angle variables, projecting down onto an action–angle plane serves to fold up all the information along the other action or angle variable. In the spheres case, with small axis separation, this can be achieved by plotting a projection of the stroboscopic map onto the action plane. We do this by taking the points of the stroboscopic map expressed in action–angle variables and forgetting about the angle coordinate; plotting the points only in the actions. One can think of the operation geometrically as concertinaing the spheres about the angle variable into one plane.

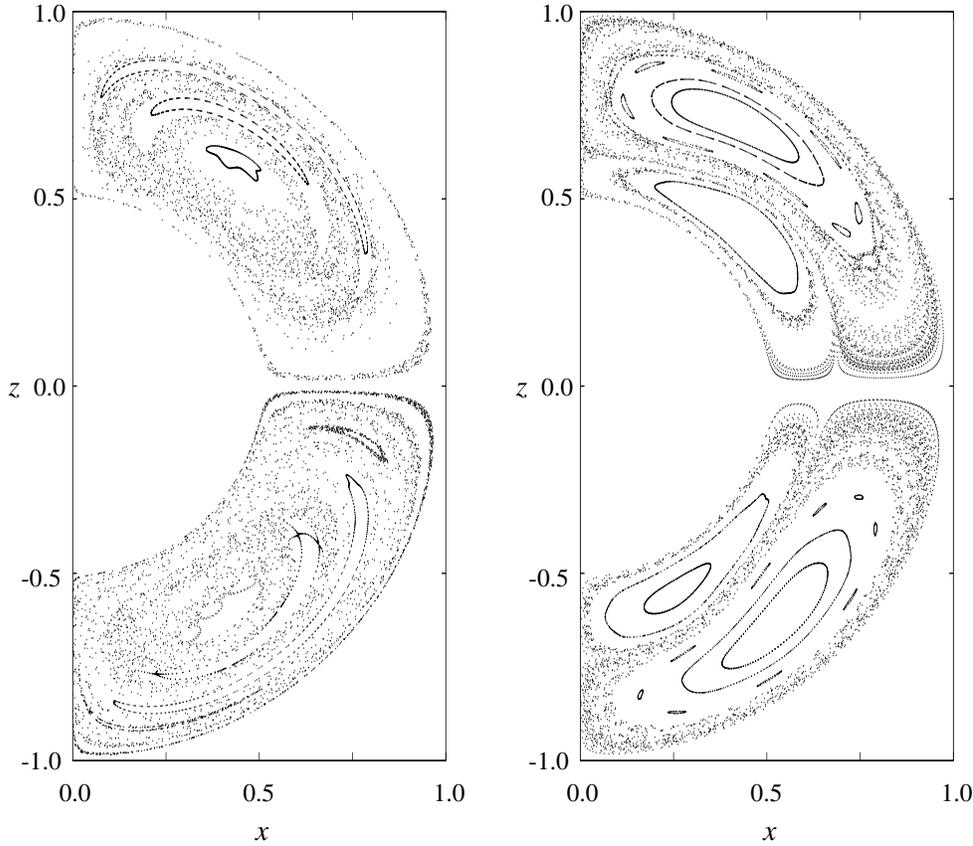

FIGURE 4. Projections of stroboscopic maps onto the plane $y = r \sin\theta \sin\phi = 0$ show Hamiltonian chaos in the flow when $\alpha = 0$. Both projections show iterates resulting from several different initial conditions. The parameter values in both cases are $\eta = 0.5$, $Re = 0.1$ and $T = 12000$. (*a*) At left: $\mu_1 = 0$, $\mu_2 = -8$, and the flow is a chaotic perturbation of two-vortex flow. (*b*) At right: $\mu_1 = 0$, $\mu_2 = -3$, and the flow is a chaotic perturbation of four-vortex flow.

Apart from the period $T$, the other new parameter in biaxial unsteady spherical Couette flow is the angle $\alpha$ between the two axes. We have chosen to have the two axes lying in the $(x, z)$-plane of our coordinate system (see figure 3), since this simplifies the algebra whilst not affecting the dynamics. If $\alpha = 0$, the spheres rotate about a single axis, and we have a $(2+1)$-dimensional flow that may be steady, if $\mu_1 = \mu_2$ — which corresponds to the spherical Couette flow described in §3.1 — or unsteady if $\mu_1 \neq \mu_2$. Whilst in the former case the flow is integrable, in unsteady spherical Couette flow the spheres rotate about a single axis with angular velocities modulated with period $T$, and the secondary flow stream function $\psi$ may be identified with a one-degree-of-freedom time-dependent Hamiltonian, which will generically exhibit chaos. Although the secondary flow in $(r, \theta)$ is decoupled from the primary flow in $\phi$, it provides input to the primary flow equation (3.17), which will be a chaotic input when the secondary flow is chaotic†. Numerically integrating the flow with $\alpha = 0$ for angular velocity ratios $\mu_1$ and $\mu_2$, where $\mu_1 \neq \mu_2$, we find the Hamiltonian chaos we expected, as we show in figure 4. We find that a significant measure of Hamiltonian chaos appears in the flow only for large values of $T$: for low-Reynolds-number flow $Re = 0.1$, the period $T$ needs to be approximately $10^4$ before Hamiltonian chaos becomes apparent in numerical simulations. The increasing amount of Hamiltonian chaos with longer periodicities has also been noted in the investigations of journal-bearing flow.

If we introduce a small coupling between $(r, \theta)$ and $\phi$, by using two rotation axes with a small angle $\alpha$ between them, we can simultaneously introduce both time dependence and fully-fledged three dimensionality to the problem. In order to analyse this, we take spherical Couette flow (3.15–3.17) and derive a Liouvillian map describing the strobed behaviour of biaxial unsteady spherical Couette flow in the limits of low Reynolds numbers and small axis separation. We have no reason to favour one of the axes over the other, so it is natural to use a coordinate system located in the $\alpha/2$-rotated reference frame $(r, \tilde{\theta}, \tilde{\phi})$ oriented midway between the two axes, that collapses back to $(r, \theta, \phi)$ when $\alpha = 0$. In this frame the map can be written

$$r' = r + \frac{Re}{2} \left(f_1(r) + f_2(r)\right) \left(3\cos^2\tilde{\theta} - 1\right) T$$
$$- \frac{3Re}{2} \left(f_1(r)\cos\tilde{\phi} - f_2(r)\left(2\cos\zeta - \cos\tilde{\phi}\right)\right) \sin\tilde{\theta} \cos\tilde{\theta}\, T\alpha + O(\alpha^2), \quad (3.19)$$

$$\tilde{\theta}' = \tilde{\theta} + \frac{Re}{2}\left(g_1(r) + g_2(r)\right) \sin\tilde{\theta}\cos\tilde{\theta}\, T$$
$$- \frac{1}{2}\left(2\cos\zeta - \cos\tilde{\phi} - \cos\xi + \frac{Re}{4}f_1(r)h_2'(r)\left(3\cos^2\tilde{\theta} - 1\right)\sin\xi\, T^2\right.$$
$$\left. - \frac{Re}{2}\left(g_1(r)\cos\tilde{\phi} - g_2(r)\left(2\cos\zeta - \cos\tilde{\phi}\right)\right)\left(2\cos^2\tilde{\theta} - 1\right)T\right)\alpha + O(\alpha^2), \quad (3.20)$$

$$\tilde{\phi}' = \tilde{\phi} + \frac{1}{2}\left(h_1(r) + h_2(r)\right) T + \frac{Re}{4}f_1(r)h_2'(r)\left(3\cos^2\tilde{\theta} - 1\right)T^2$$
$$+ \frac{1}{2}\left(2\operatorname{cosec}\zeta - \operatorname{cosec}\tilde{\phi} - \operatorname{cosec}\xi + \frac{Re}{2}\left(\left(g_1(r) + g_2(r)\right)\operatorname{cosec}\xi - 2g_1(r)\operatorname{cosec}\zeta\right.\right.$$
$$\left.\left. + \frac{1}{2}f_1(r)h_2'(r)\left(\left(3\cos^2\tilde{\theta} - 1\right)\operatorname{cosec}\xi\cot\xi - 6\sin^2\tilde{\theta}\cos\tilde{\phi}\right)T\right)T\right)\cot\tilde{\theta}\,\alpha$$
$$+ O(\alpha^2), \quad (3.21)$$

where $h_2'(r) = \mathrm{d}h_2(r)/\mathrm{d}r$,

$$\zeta = \tilde{\phi} + \frac{1}{2}h_1(r)T, \qquad \xi = \tilde{\phi} + \frac{1}{2}\left(h_1(r) + h_2(r)\right)T, \quad (3.22)$$

† Thus in order to fully understand transport in $(2+1)$-dimensional flows like spherical Couette flow, it is not enough to consider the Hamiltonian dynamics of the secondary flow alone. Nevertheless, it is possible to determine many properties of the flow from the secondary flow dynamics, and since the secondary flow equations (3.15) and (3.16) have the Reynolds number *Re* as an overall multiplying factor, if one is only interested in the secondary flow dynamics, the product *Re T* can be treated as a single parameter.

and

$$f_j(r) = \frac{1}{r^2}\left(\frac{A_{1j}}{r^2} + A_{2j} + A_{3j}r^3 + A_{4j}r^5 + \frac{a_{2j}}{4}\left(\frac{a_{2j}}{r} - a_{1j}r^2\right)\right), \quad (3.23)$$

$$g_j(r) = -\frac{1}{r^2}\left(-\frac{2A_{1j}}{r^3} + 3A_{3j}r^2 + 5A_{4j}r^4 + \frac{a_{2j}}{4}\left(-\frac{a_{2j}}{r^2} - 2a_{1j}r\right)\right), \quad (3.24)$$

$$h_j(r) = a_{1j} + \frac{a_{2j}}{r^3}, \quad (3.25)$$

and $j$ is 1 or 2; $a_{ij}$ and $A_{ij}$ being $a_i$ (3.6) and $A_i$ (3.7–3.10), respectively, evaluated with $\mu = \mu_j$. The terms of order $\alpha$ are those through which $r$ and $\tilde{\theta}$ couple to $\tilde{\phi}$ and perturb the $(2+1)$-dimensional flow into being fully three dimensional. Note that when $\mu_1 = \mu_2$, so that $f_1(r) = f_2(r)$, $g_1(r) = g_2(r)$, and $h_1(r) = h_2(r)$, the two semiperiods are identical and there is cancellation within the the order-$\alpha$ terms which results in the coupling being weaker than otherwise.

When $Re$ and $\alpha$ are small, The map (3.19–3.21) has the same form as the perturbed two-action Liouvillian map (2.4), with $r$ and $\tilde{\theta}$ as the actions and $\tilde{\phi}$ as the angle variable. As we showed in §2, a map of the action plane $(r, \tilde{\theta})$ can be constructed by averaging (3.19–3.20) over the angle $\tilde{\phi}$, giving

$$r' = r + \frac{Re}{2}(f_1(r) + f_2(r))\left(3\cos^2\tilde{\theta} - 1\right)T, \quad (3.26)$$

$$\tilde{\theta}' = \tilde{\theta} + \frac{Re}{2}(g_1(r) + g_2(r))\sin\tilde{\theta}\cos\tilde{\theta}\,T. \quad (3.27)$$

Since, as we mentioned above, $Re\,T$ can be treated as a single parameter for the purposes of the secondary flow, taking either of the limits $Re \to 0$ or $T \to 0$ of the Euler map (3.26–3.27) has the same effect, and the actions evolve according to the stream function

$$H = \frac{1}{2}r^2(f_1(r) + f_2(r))\sin^2\tilde{\theta}\cos\tilde{\theta}, \quad (3.28)$$

$$= \frac{1}{2}\left(\frac{A_{11} + A_{12}}{r^2} + (A_{21} + A_{22}) + (A_{31} + A_{32})r^3 + (A_{41} + A_{42})r^5 \right.$$

$$\left. + \frac{a_{21}}{4}\left(\frac{a_{21}}{r} - a_{11}r^2\right) + \frac{a_{22}}{4}\left(\frac{a_{22}}{r} - a_{12}r^2\right)\right)\sin^2\tilde{\theta}\cos\tilde{\theta}. \quad (3.29)$$

The invariant tori in the small-axis-separation limit are then the level surfaces of (3.29). Note that (3.29) reduces to (3.4), up to a constant, when $\mu_1 = \mu_2$. The resonant surfaces in the small-axis-separation limit may be obtained as in (2.8) by considering the values of the action variables for which the above averaging procedure fails, when the angle variable changes by a rational fraction of $2\pi$ each period. We consider (3.21) in the limits $Re \to 0$ and $\alpha \to 0$, so that the terms of order $\alpha$ and $Re$ disappear and

$$\tilde{\phi}' - \tilde{\phi} = \frac{2\pi k}{n} = \frac{1}{2}(h_1(r) + h_2(r))\,T. \quad (3.30)$$

The resonances are then the solutions of (3.30) where $k$ and $n$ are integers. These occur when

$$r_{k/n} = \left(\frac{(1-\bar{\mu})\eta^3 nT}{(1-\bar{\mu}\eta^3)nT - 2\pi k}\right)^{\frac{1}{3}}, \quad (3.31)$$

which are shells of constant $r$, whose radius depends on the mean value of $\mu_1$ and $\mu_2$, $\bar{\mu} = (\mu_1 + \mu_2)/2$, as well as on $\eta$, $T$ and the order $k/n$ of the resonance. For example, in figure 5 (*a*) we show the distribution of some of the resonances in a two-vortex flow with $\eta = 1/2$, $\mu_1 = 0$, $\mu_2 = -8$, and $T = 4$. The terms of order $\alpha$ and $Re$ in (3.21) constitute negligible perturbations of these resonance shells when the axis separation and the Reynolds number are small. The resonances are dense throughout space, but they have a hierarchical structure in which their relative strengths are governed by the Fourier expansion of (3.19–3.21), such that in general the resonances with smaller denominators $n$ are stronger; the primary resonances where $n = 1$ have the largest influence, and the effect of the secondary and higher-order resonances $n \geqslant 2$ is far smaller. One can tune the resonances using $\bar{\mu}$ and $T$ so that more or less primary resonances be inside or outside the region $\eta < r < 1$ between the spheres. In figure 5 (*b*) we show an example with $\eta = 1/2$, $\mu_1 = 0$, $\mu_2 = -3$, and $T = 20$ of a four-vortex flow with more primary resonances than figure 5 (*a*).

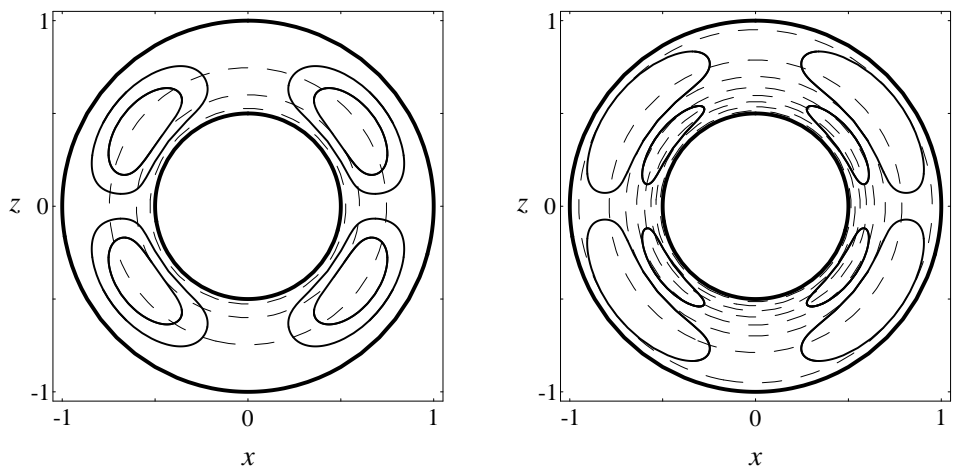

FIGURE 5. Primary resonances (3.31) — shown dashed — and adiabatic invariant surfaces (3.29) in biaxial spherical Couette flow are shown for small axis separation in the low-Reynolds-number regime. (*a*) At left: $\eta = 0.5$, $\mu_1 = 0$, $\mu_2 = -8$, and $T = 4$; a two-vortex flow. The resonances in the spheres are shown where, from the inner sphere outwards, $k/n$ is $-2/1$, $-1/1$, and $0/1$. (*b*) At right: $\eta = 0.5$, $\mu_1 = 0$, $\mu_2 = -3$, and $T = 20$; a four-vortex flow. The increase in the axis-switching period $T$ from (*a*) means there are more resonances in this the picture: those where $k/n$ is $-4/1$, $-3/1$, $-2/1$, $-1/1$, $0/1$, $1/1$, $2/1$, and $3/1$ are shown.

In the projection of the stroboscopic map of the flow in figure 6 (*a*), we can see that the trajectory is, as predicted, lying on the invariant tori of (3.29), except for intervals where it jumps from one invariant torus to another. Also shown dashed in figure 6 (*a*) is the principal resonant surface, and it is apparent that the resonances are the means by which the trajectory jumps from invariant torus to invariant torus. To compare with figure 6 (*a*), in figure 6 (*b*) we show an example of the same resonance-induced dispersion behaviour occurring in a two-action Liouvillian map. As we increase the coupling $\alpha$, the resonances have increasing influence, and the length of time for which trajectories are captured into resonance increases. In figure 7 (*a*) we show another projection of a stroboscopic map, this time for a larger value of $\alpha$ than in figure 6 (*a*), but with the other parameters remaining unchanged. Trajectories are now captured by resonances for much longer times, and by this means they can rapidly cross the separatrix between the vortices at the equator and disperse throughout practically the whole region between the spheres. This is shown graphically in figure 8 (*a*), which illustrates a slice between $-0.01 < y < 0.01$ through the same stroboscopic map as figure 7 (*a*). The stroboscopic map slice shows the evolution of a single trajectory, and comparing it with the distribution of the resonances shown in figure

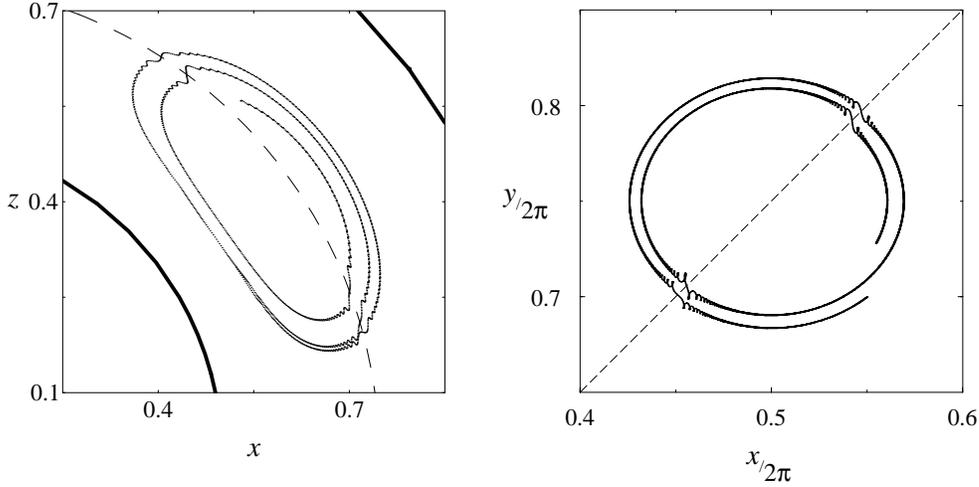

FIGURE 6. (*a*) At left: a projection of a stroboscopic map onto the plane $y = r \sin \tilde{\theta} \sin \tilde{\phi} = 0$ illustrates the jump behaviour typical of resonance-induced dispersion in two-action flows. The parameter values are as in figure 5 (*a*), with $Re = 0.1$ and $\alpha = 0.1°$. These parameter values are such as to give a two-vortex flow, and to put the primary 0/1 resonance in the middle of the region between the spheres, whilst leaving other primary resonances distant. Notice that because this is a projection, trajectories can appear to cross. (*b*) At right: a projection of a trajectory in a Liouvillian map shows resonance-induced dispersion. The diagonal dashed line indicates the location of the primary resonance. Close to this line the trajectory oscillates wildly and jumps from one invariant torus to another. The map used is $x' = x + 0.001(\sin z + 2\cos y)$, $y' = y + 0.001(1.5 \sin x' + 2.5 \cos z)$, $z' = z + 4(\cos y' + \sin x')$.

5 (*a*), we observe that the only regions as yet unvisited in figure 8 (*a*) are close to the poles, and the centres of the vortices that lie inside the resonance 0/1 shell and are not crossed by any primary resonance. Resonance-induced dispersion into these regions will occur slowly through weak high-order resonances that lie densely between the primary resonances. To compare with this global mixing behaviour in the two-vortex case, in figures 7 (*b*) and 8 (*b*) we show that for four-vortex flow, resonance-induced dispersion operates separately within the inner and outer vortices; there is no mixing across the separatrix between them, since this separatrix is concentric with, and therefore not crossed by, any of the resonance shells.

Looking at figure 7 side by side with the $\alpha = 0$ Hamiltonian chaos of figure 4 shows how different resonance-induced dispersion appears in comparison; the absence of KAM tori shows that we are dealing with chaos that is not a small perturbation of an integrable Hamiltonian system. Whilst the Hamiltonian chaos of figure 4 is produced by long-period flow, $T = 12000$, the resonance-induced dispersion of figure 7 is already present when the flow period is much shorter; $T = 4$. At these short periodicities, the measure of Hamiltonian chaos is minute. In the case where $\mu_1 = \mu_2$ there is no Hamiltonian chaos when $\alpha = 0$. Recall from our discussion of (3.19–3.21) that when $\mu_1 = \mu_2$ the coupling between $(r, \tilde{\theta})$ and $\tilde{\phi}$ is weaker than otherwise. This implies that when $\alpha \neq 0$ and $\mu_1 = \mu_2$ all the resonances will have less influence than normal. Furthermore, there is additional cancellation that takes place for $r = r_{0/1}$ when $\mu_1 = \mu_2$; in this case $h_1(r) = h_2(r) = 0$ and the order-$\alpha$ terms cancel completely, so that the 0/1 resonance is not present when $\mu_1 = \mu_2$. This anomalous behaviour when $\mu_1 = \mu_2$ is why we have chosen throughout this paper to use examples where $\mu_1 \neq \mu_2$ for illustrating resonance-induced dispersion.

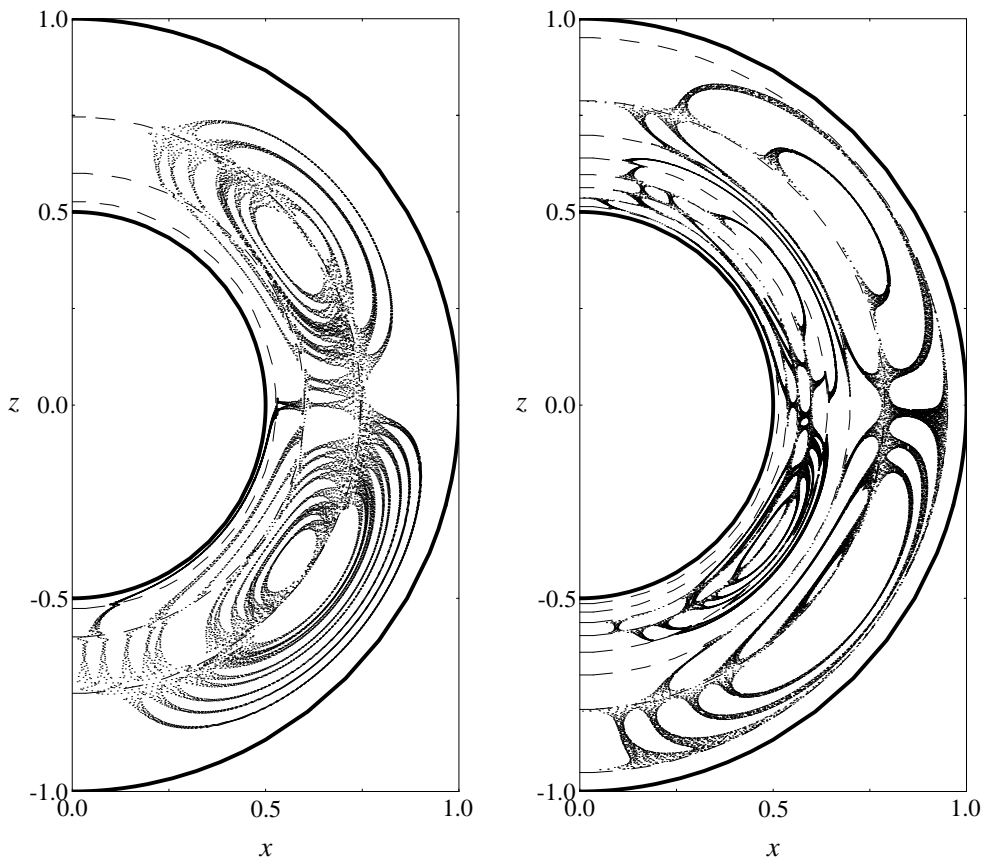

FIGURE 7. Projections of stroboscopic maps onto the plane $y = r\sin\tilde\theta\sin\tilde\phi = 0$ for $Re = 0.1$ and $\alpha = 1°$. (*a*) At left: other parameters are as in figure 5 (*a*), and a single trajectory is depicted in this two-vortex flow. (*b*): At right: other parameters are as in figure 5 (*b*), and two trajectories are shown. This is a four-vortex flow; there is no mixing across the separatrix between the inner and outer vortices.

There are several parameters one can alter to control the rate of resonance-induced dispersion. Tuning the resonances with (3.31) is one option. The stroboscopic map projections of figures 6 (*a*) and 7 (*a*) were made with $\mu_1 = 0$ and $\mu_2 = -8$ so that the 0/1 primary resonance passed right through the centre of the region $\eta < r < 1$ and cut almost all invariant tori of (3.29), to ensure that resonance-induced dispersion with a primary resonance would occur every time the trajectory wound once around the vortex. Increasing the axis-switching period $T$ would speed up the dispersion by increasing the density of primary resonances, as we showed in figure 5 (*b*). If, on the other hand, we choose to tune the system by changing $\bar\mu$ and $T$ so that all the primary resonances lie outside the interval $\eta < r < 1$, we would still have resonance-induced dispersion, but operating only with the weaker secondary and higher resonances, so that the dispersion would be slower with smaller jumps. Following the discussion above, another possibility is to make $\mu_1$ the same as or different to $\mu_2$, as the resonances are stronger when $\mu_1 \neq \mu_2$. We can also alter the axis separation $\alpha$, since the resonances gain strength with increasing $\alpha$, as the comparison of figure 6 (*a*) with figure 7 (*a*) shows.

For a different view of resonance-induced dispersion, we can look at the time evolution of the stream function (3.29). If the adiabatic approximation of (3.23–3.25) were exact, this would be constant in time. Instead, if resonance-induced dispersion were occurring, we would expect to see that this graph should be almost flat when the trajectory evolved on a level surface of an invariant torus, but these flat regions would be interspersed with oscillations leading to sudden jumps, as each time the trajectory approached a resonance intersecting the invariant torus it would be transported to another level surface with a different value of $H$, rather like playing a game of snakes and ladders. Figure 9 (*a*), for the same parameter values and initial conditions as figure 6 (*a*), is such a graph, and indeed we observe this snakes-and-ladders behaviour exactly as we have just described — we are seeing resonance-induced dispersion in action. As an example to compare with figure 9 (*a*), we show in figure 9 (*b*) a similar graph of resonance-induced dispersion in a two-action Liouvillian map.

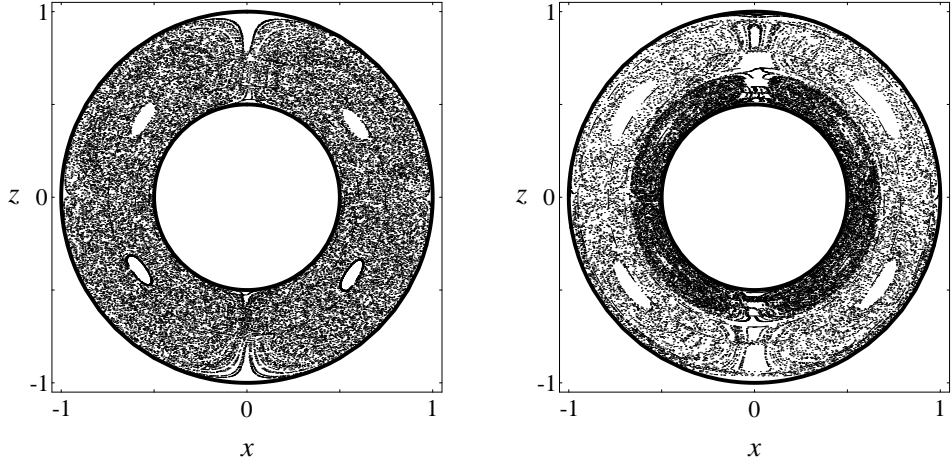

FIGURE 8. Slices of stroboscopic maps between $-0.01 < y < 0.01$. (*a*) At left: the parameter values and initial conditions are as in figure 7 (*a*); the resonances are distributed as in figure 5 (*a*). Forty thousand points (corresponding to many times this number of periods) are shown, all from the same initial condition. This slice illustrates how effective resonance-induced dispersion is at mixing; the only regions where the trajectory has not yet ventured are close to the poles, and the centres of the vortices, which lie inside the $0/1$ resonance. In time, the trajectory will disperse into these regions too, through the action of the higher-order resonances that are dense throughout the space. If, by altering $\bar{\mu}$, the $0/1$ resonance were tuned to pass right through the centres of the vortices, these regions would be visited as frequently as others. (*b*) At right: two initial conditions are used to illustrate dispersion in the trajectory with parameter values as in 7 (*b*); the resonances are distributed as in figure 5 (*b*). The space is divided into two regions by the separatrix between the two vortices in each hemisphere. There in no mixing between these two regions, but effective mixing occurs separately within each region.

We find numerically that the small-axis-separation approximation for biaxial unsteady spherical Couette flow remains good as long as $\alpha$ is less than around ten degrees. The calculations of invariant tori and resonant surfaces we made for the small-axis-separation flow become increasingly poor approximations when the axis separation $\alpha$ is made larger. For $\alpha$ greater than one degree, the noise arising from the fact that the small-axis-separation approximation is just an approximation starts to overwhelm the signal. However, we can recuperate it by applying a running average to the noisy signal from (3.29), as we show in figure 10. For $\alpha$ more than about ten degrees, this averaging starts to lose its effectiveness, and consequently we are unable perform much analysis for larger axis separations. We believe that the large-axis-separation flow is still of two-action type; the interaction between $\tilde{\theta}$ and $\tilde{\phi}$ will produce one action and one angle variable, $r$ being the other action. However since the large-axis-separation flow is not a small perturbation of a $(2+1)$-dimensional flow, we are unable to obtain analytical expressions for the action and angle variables in the large-axis-separation case in the same way as was possible for small-axis-separation flow.

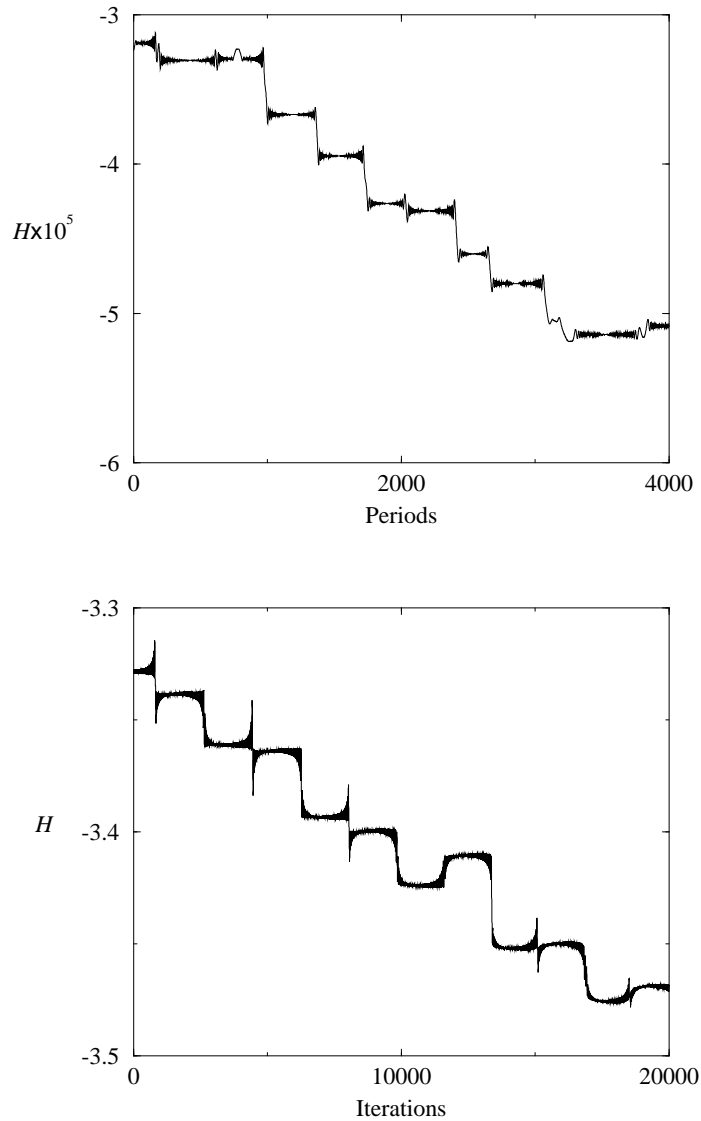

FIGURE 9. (*a*) Above: resonance-induced dispersion is illustrated here with a graph of the adiabatic invariant (3.29) against time of the trajectory in figure 6 (*a*). (*b*) Below: resonance-induced dispersion in a Liouvillian map is shown using a graph of the adiabatic invariant (2.7) against time of the trajectory in figure 6 (*b*). In both (*a*) and (*b*) the adiabatic invariant remains nearly constant almost all the time, but jumps chaotically from one invariant torus to another at each intersection with a primary resonance. The general downward trend in the graphs corresponds to the trajectory winding towards a vortex centre, as is visible in figures 6 (*a*) and (*b*).

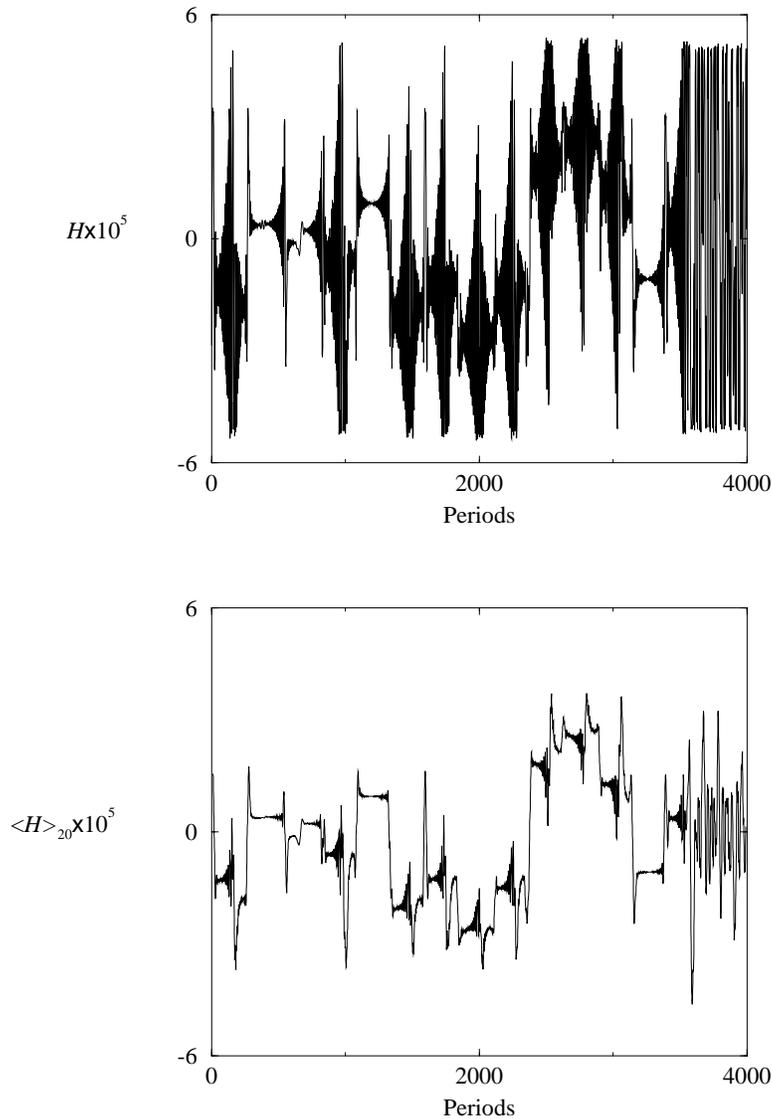

FIGURE 10. Averaging at work; the initial conditions and parameter values here are as in figure 6 (*a*), except for the axis separation angle $\alpha$, which is now $10°$. The upper graph shows the noisy adiabatic invariant $H$ (3.29) against time. The noise arises because $H$ is only approximately the invariant, and the approximation worsens with increasing $\alpha$. In the lower graph we have performed a twenty-point running average on the trajectory. The running average has smoothed out the noise substantially, so that we can now see the jumps much more clearly. This shows that even in cases where we do not possess the exact adiabatic invariant, it is still possible using averaging to see the characteristic two-action jump behaviour at work.

## 4. Chaotic advection in three-dimensional unsteady flows

Following our classification in §2 of three-dimensional unsteady flows by the number of fast angle and slow action variables, biaxial unsteady spherical Couette flow is of the two-action class. Three-dimensional flows often arise from what we have termed $(2+1)$-dimensional flows by the coupling of a one-dimensional primary flow to a two-dimensional secondary flow, giving one variable of one type and two variables of the other. Generically this will preclude zero-action and three-action flows, but a large class of three-dimensional unsteady flows are two-action flows, and one-action flows will be equally common (Mezić & Wiggins 1994). With biaxial unsteady spherical Couette flow the relationship between the primary and secondary flow rates depends on the Reynolds number — the secondary flow is hundreds to thousands of times slower than the primary flow when $Re = 0.1$ — so that although it would be possible in our model to increase the Reynolds number sufficiently to give two fast secondary variables and one slow primary variable, making the flow of one-action type, doing so would make the low-Reynolds-number approximation we started with invalid, landing us squarely in the turbulent regime (Bühler 1992), Other possible candidate flows for obtaining a one-action three-dimensional unsteady flow model that can both be treated analytically and is experimentally realizable should be sought out and investigated.

We have seen in biaxial unsteady spherical Couette flow that the zero-axis-separation limit of $\alpha = 0$ can lead us either to two-dimensional steady flow without chaos, if $\mu_1 = \mu_2$, or to two-dimensional unsteady flow with chaos, if $\mu_1 \neq \mu_2$. The limit we cannot explore with biaxial unsteady spherical Couette flow corresponds to the transition between three-dimensional unsteady flow and three-dimensional steady flow. One way to arrive at a three-dimensional steady flow model using spherical Couette flow would be to displace the spheres so that they would no longer be coaxial, like the cylinders in journal-bearing flow. In this way we would break the $\phi$ symmetry and obtain a three-dimensional steady flow, spherical-bearing flow, which could then be modulated in time using different $\mu_1$ and $\mu_2$ to give a three-dimensional unsteady flow. The only work we know of on this geometry is that of Wannier (1950), who however only treats the the problem of spherical-bearing flow in the limit of a very small gap between the spheres, such that the radial component of the motion can be assumed to be zero. This limit of the Stokes equation leads to the Reynolds equations of lubrication theory, which give a two-dimensional flow in $\theta$ and $\phi$. Munson (1974) solves perturbatively the case where the spheres are displaced along their common axis, so that they are eccentric but remain coaxial. This however retains the $\phi$ symmetry, so it is another example of a $(2 + 1)$-dimensional steady flow rather than a three-dimensional steady flow.

Another possibility for investigating the transition between three-dimensional steady and unsteady flow is given by the flow introduced by Bajer & Moffatt (1992), who treat the three-dimensional steady flow between two concentric spheres rotating about different axes. This flow is particularly interesting because it exhibits the consequences of the breakdown of the Hamiltonian description of steady three-dimensional flows that occurs at stagnation points of the flow. This phenomenon is similar in origin to the resonance-induced dispersion we have been investigating here in three-dimensional unsteady flows. In both cases averaging fails in certain regions of space. In three-dimensional unsteady flows these regions are generically two-dimensional resonant surfaces, whereas in three-dimensional steady flows they are zero-dimensional stagnation points. However, despite the less important rôle, occasioned by their lower dimensionality, that these regions play in three-dimensional steady flow, in appropriate circumstances the global topology of the flow can be altered by the breakdown of averaging at the stagnation points, leading to what Bajer & Moffatt have termed trans-adiabatic drift (Bajer & Moffatt 1990; Bajer 1994). Trans-adiabatic drift can be considered to be the equivalent in three-dimensional steady flows of resonance-induced dispersion. It should thus be particularly interesting to model the addition of time dependence to this flow. We should point out that adiabatic descriptions of processes that are valid everywhere except in certain neighbourhoods are ubiquitous in many areas of physics, and lead to graphs of the type of figure 9 being seen in many contexts.

Occupying as they do a position somewhere between symplectic maps of two and four dimensions from two and three-degree-of-freedom Hamiltonian systems, the Liouvillian maps that underlie three-dimensional unsteady flows can exhibit types of behaviour analogous to both their lower and higher-dimensional siblings. An important difference between Hamiltonians of two and three degrees of freedom is that Arnold diffusion† exists in the latter but not in the former. With two degrees of freedom, the invariant KAM tori completely separate different regions of phase space. This is no longer true when another degree of freedom is added, and as a result Arnold diffusion can take place throughout the phase space. This occurs, albeit extremely slowly, for arbitrarily small values of the nonlinearity parameter $\varepsilon$. Resonance-induced dispersion in two-action Liouvillian maps, although reminiscent of Arnold diffusion in the four-dimensional volume-preserving maps from three-degree-of-freedom Hamiltonians, is engendered by a different mechanism. Whilst Arnold diffusion occurs through an interconnected web of resonances, in resonance-induced dispersion the available space is covered using the invariant tori, and the resonant surfaces are a means to jump from one torus to another. The consequence of this difference is that the rates of the two processes are radically different; with nonlinearity $\varepsilon$, an estimate of the upper bound on the rate of Arnold diffusion is $O(\exp(-\varepsilon^{-1/2}))$, whilst previous work on Liouvillian maps shows that resonance-induced dispersion has an $O(\varepsilon^2)$ dependence (Piro & Feingold 1988).

　　The presence of resonance-induced dispersion in two-action type flows, and its absence in those of one-action type, is from the mixing and transport point of view the crucial difference between these two classes of flows. In the one-action class of three-dimensional unsteady flows, as with two-dimensional unsteady flows, the existence of KAM-type theorems proves that complete mixing will not generically be achievable, as the invariant tori present barriers to transport. On the other hand, the addition of a third dimension to a two-dimensional unsteady flow leads in the two-action case to much faster dispersion than with two-dimensional flow alone, owing to the presence of resonance-induced dispersion. This should lead to better mixing. What we mean by efficient mixing here is that the density of an initial concentration of passive scalars should rapidly tend to a constant value throughout the region being mixed. This property is related to the Lyapunov exponents of the flow, one of which must be positive to allow the exponential separation of neighbouring trajectories that is necessary for chaos. We shall be reporting in a future work investigations of Lyapunov exponents and mixing efficiencies in biaxial unsteady spherical Couette flow and other three-dimensional unsteady flows. Even a small coupling between primary and secondary flows is effective in two-action flows in producing a large improvement in dispersion over two-dimensional chaotic advection alone through the action of resonance-induced dispersion, particularly when the flow period is relatively short when there is very little chaotic advection in two dimensions. In practical terms, we have seen that even a wobble from a bad bearing between the two spheres leading to a free play of $0.1°$ is enough to produce a spectacular growth in dispersion rate. Although we have run some of our plots for long times, by adjusting the parameters it is possible to arrive at a flow in which the effects of resonance-induced dispersion are visible after only a handful of periods.

　　Biaxial unsteady spherical Couette flow forms a bridge between theory and experiment for investigating chaotic advection in three-dimensional unsteady flows; we have an analytical solution for the stream function and the flow can be realized in the laboratory. Journal-bearing flow has performed this same function extremely well for two-dimensional unsteady flows. We hope that the model we have presented here might spur similar interest in the analysis and investigation of chaotic advection in three-dimensional unsteady flows, and that the challenge of devising suitable apparatus for the experimental observation of resonance-induced dispersion and the other phenomena described here will be taken up.

---

† One of the terminological differences between nonlinear dynamics and fluid dynamics is that *diffusion* in nonlinear dynamics refers to a process that in fluid dynamics is called *dispersion* (diffusion in fluid dynamics carrying connotations of molecular diffusion). In previous papers (Piro & Feingold 1988; Feingold, Kadanoff & Piro 1989; Cartwright, Feingold & Piro 1994b, 1995) we used the term diffusion in the dynamical systems sense.

We should like to thank Konrad Bajer, Grae Worster, and an anonymous referee for their useful comments and suggestions about a previous version of this work. MF acknowledges the support of the Israel Science Foundation administered by the Israel Academy of Sciences and Humanities, and OP and JHEC that of the Spanish Dirección General de Investigación Científica y Técnica, contract number PB92-0046-c02-02 and European Union Human Capital and Mobility contract number ERBCHBICT920200.